\documentclass[prc,twocolumn,showpacs,floatfix,nofootinbib,preprintnumbers,superscriptaddress,amsmath,amssymb]{revtex4}

\usepackage{amsmath}           
\usepackage{amsfonts}
\usepackage{graphicx}          
\usepackage{dcolumn}           
\usepackage{bm}

\usepackage{color}

\begin{document}

\bibliographystyle{apsrev}


\title{Nuclear energy density optimization: \\Large deformations}

\author{M. Kortelainen}
\affiliation{Department of Physics and Astronomy, University of Tennessee, Knoxville, TN 37996, USA}
\affiliation{Physics Division, Oak Ridge National Laboratory, P.O. Box 2008, Oak Ridge, TN 37831, USA}

\author{J. McDonnell}
\affiliation{Department of Physics and Astronomy, University of Tennessee, Knoxville, TN 37996, USA}
\affiliation{Physics Division, Oak Ridge National Laboratory, P.O. Box 2008, Oak Ridge, TN 37831, USA}

\author{W. Nazarewicz}
\affiliation{Department of Physics and Astronomy, University of Tennessee, Knoxville, TN 37996, USA}
\affiliation{Physics Division, Oak Ridge National Laboratory, P.O. Box 2008, Oak Ridge, TN 37831, USA}
\affiliation{Institute of Theoretical Physics, Warsaw University, ul. Ho\.{z}a 69, PL-00681, Warsaw, Poland}

\author{P.-G. Reinhard}
\affiliation{Institut f\"ur Theoretische Physik, Universit\"at Erlangen, D-91054 Erlangen, Germany}

\author{J. Sarich}
\affiliation{Mathematics and Computer Science Division, Argonne National Laboratory, Argonne, IL 60439, USA}

\author{N. Schunck}
\affiliation{Physics Division, Lawrence Livermore National Laboratory, Livermore, CA 94551, USA}
\affiliation{Department of Physics and Astronomy, University of Tennessee, Knoxville, TN 37996, USA}
\affiliation{Physics Division, Oak Ridge National Laboratory, P.O. Box 2008, Oak Ridge, TN 37831, USA}

\author{M. V. Stoitsov}
\affiliation{Department of Physics and Astronomy, University of Tennessee, Knoxville, TN 37996, USA}
\affiliation{Physics Division, Oak Ridge National Laboratory, P.O. Box 2008, Oak Ridge, TN 37831, USA}

\author{S. M. Wild}
\affiliation{Mathematics and Computer Science Division, Argonne National Laboratory, Argonne, IL 60439, USA}


\newcommand{\cS}{\mathcal{S}}	
\newcommand{\cN} {\mbox{$\cal N$}} 
\newcommand{\Ref}[1]{\mbox{\rm{(\ref{#1})}}}    
\newcommand{\qn}{\frac{(n+1)(n+2)}{2}}   
\newcommand{\R}{\mbox{${\mathbb R}$}}           
\newcommand{\xb}{\mathbf{x}}  
\newcommand{\cb}{\mathbf{c}}  
\newcommand{\xh}{\hat{\xb}}  
\newcommand{\nx}{n_x}  
\newcommand{\iset}{\mathcal{X}} 
\newcommand{\tb}{\bm{\theta}} 
\newcommand{\nt}{n_{\theta}}  
\newcommand{\nd}{n_d}  
\newcommand{\eps}{\varepsilon} 
\newcommand{\epsb}{\bm{\eps}} 
\newcommand{\gb}{\mathbf{g}}  
\newcommand{\Hb}{\mathbf{H}}  
\newcommand{\cB} {\mbox{$\cal B$}} 
\newcommand{\Fb}{\mathbf{F}}  
\newcommand{\db}{\bm{\delta}} 
\newcommand{\cov}{\text{Cov}}
\newcommand{\var}{\text{Var}}
\newcommand{\algo}{\textsc{pound}er\textsc{s}}

\newcommand{\BdG}{\textsc{b}{\footnotesize d}\textsc{g}}
\newcommand{\PWscf}{\textsc{pw}scf}
\newcommand{\DFT}{\textsc{dft}}
\newcommand{\RMF}{\textsc{rmf}}
\newcommand{\SLDA}{\textsc{slda}}
\newcommand{\HO}{\textsc{ho}}
\newcommand{\THO}{\textsc{tho}}
\newcommand{\HF}{\textsc{hf}}
\newcommand{\BCS}{\textsc{bcs}}
\newcommand{\LN}{\textsc{ln}}
\newcommand{\EFA}{\textsc{efa}}
\newcommand{\HFODD}{\textsc{hfodd}}
\newcommand{\HFBTHO}{\textsc{hfbtho}}
\newcommand{\CC}{\textsc{cc}}
\newcommand{\CCSD}{\textsc{ccsd}}
\newcommand{\BLAS}{\textsc{blas}}
\newcommand{\LAPACK}{\textsc{lapack}}
\newcommand{\ATLAS}{\textsc{atlas}}
\newcommand{\GNU}{\textsc{gnu}}
\newcommand{\CECO}{\textsc{ceco}}
\newcommand{\UNEDFZERO}{\textsc{unedf0}}
\newcommand{\UNEDFNB}{\textsc{unedf}nb}
\newcommand{\exclude}[1]{}
\newcommand{\mat}[1]{\mathbf{#1}}
\newcommand{\ket}[1]{|#1\rangle}
\newcommand{\bra}[1]{\langle#1|}
\newcommand{\norm}[1]{\lVert{#1}\rVert}
\newcommand{\braket}[1]{\mathinner{\langle{#1}\rangle}}{\catcode`\|=\active
  \gdef\Braket#1{\left<\mathcode`\|"8000\let|\bravert {#1}\right>}}
\newcommand{\bravert}{\egroup\,\vrule\,\bgroup}

\newcommand{\Ck}{C_\text{k}}
\newcommand{\enm}{E^\text{NM}}
\newcommand{\rhoc}{\rho_\text{c}}
\newcommand{\knm}{K^\text{NM}}

\newcommand{\asym}{a_\text{sym}^\text{NM}}
\newcommand{\lsym}{L_\text{sym}^\text{NM}}
\newcommand{\ksym}{\Delta K^\text{NM}}

\newcommand{\gras}[1]{\boldsymbol{#1}}
\newcommand{\capital}[1]{\mathscr{#1}}

\newcommand{\ba}{\begin{array}}
\newcommand{\ea}{\end{array}}
\newcommand{\disregard}[1]{}

\newcommand{\ali}{ali}

\newcommand{\UNEDFONE}{\textsc{unedf1}}
\newcommand{\UNEDF}{\textsc{unedf}}
\newcommand{\cexpar}{x^{\rm Coul}_{\rm Exc}}

\begin{abstract}
A new Skyrme-like energy density suitable for studies of strongly elongated nuclei has been determined in the framework of the Hartree-Fock-Bogoliubov 
theory using the recently developed  model-based, derivative-free optimization algorithm {\algo}. A sensitivity analysis at the optimal solution has revealed the importance of states at large deformations in driving the parameterization of the functional. The good agreement with experimental data on masses and separation energies, achieved with the previous parameterization {\UNEDFZERO}, is largely preserved. In addition, the new energy density {\UNEDFONE} gives a much improved description of the fission barriers in $^{240}$Pu and neighboring nuclei.
\end{abstract}


\maketitle


\section{Introduction}
\label{Sec-introduction}

One of the focus areas of the UNEDF SciDAC collaboration \cite{[Ber07],[Fur11]} has been 
the description of the fission process within a self-consistent framework based on 
nuclear density functional theory (DFT). 
Until now, attempts at going beyond the macroscopic-microscopic methods 
\cite{[Mol01]} have been carried in the context of the original self-consistent 
nuclear mean-field theory \cite{[Ben03]} with  Skyrme 
 (see, e.g., Refs.~\cite{[Bur04],[Bon04],[Sam05],[Sta09],[She09],[Sch11a]}), Gogny  
(see, e.g., Refs.~\cite{[War02],[Gou05],[Dub08],[You09]}),
and relativistic  (see, e.g., Refs.~\cite{[Bur04],[Rut95a],[Abu10]}) energy density functionals (EDFs).
The fundamental assumption of the nuclear DFT is that one can describe 
accurately a broad range of phenomena in nuclei, including excited states and 
large-amplitude collective motion, by enriching the density dependence of the 
functional while staying at the single-reference 
Hartree-Fock-Bogoliubov (HFB) level. In this picture, 
beyond-mean-field corrections are implicitly built-in. Preliminary studies 
aimed at re-examining the old problem of restoring broken symmetries in this 
context are promising \cite{[Dob09g],[Hup11]}. 

A common challenge to both the self-consistent mean field and the DFT approach 
is the need to carefully optimize the EDF parameters   to the preselected pool of observables \cite{[Ben03],[Ber05],[Nik08a],[Kor08],[Gor09],[Klu09],[Rei10],[Kor10]}. 
In particular, special attention must be paid to optimize the parameters in the 
same regime where the theory will later be applied and to choose the fit observables  accordingly. In a recent work \cite{[Nik10]}, we showed that existing Skyrme EDFs exhibit a significant spread in 
bulk deformation properties, and re-emphasized \cite{[Ton84],[Gue80]}
that the resulting 
theoretical uncertainties could be greatly reduced by considering data corresponding to large 
deformation in the optimization process. Let us recall that the 
early  Skyrme-type EDF SkM$^\ast$ \cite{[Bar82]} was in fact optimized for fission studies in the 
actinide region by considering
the experimental information on the fission barrier of $^{240}$Pu. However, the optimization was not performed directly at the 
deformed HFB level; instead, a semi-classical approach was used based on the Thomas-Fermi 
approximation together with shell-correction techniques. The D1S parameterization 
of the finite-range Gogny force was also fine-tuned to the first barrier height of $^{240}$Pu  \cite{[Ber89]}, considering a rotational correction to the energy of the deformed state. However,  this fine-tuning again was not done directly 
at the HFB level but by a manual readjustment of the surface coefficient of the EDF using a 
phenomenological model. Also, in the Bsk14
EDF of the HFB-14 mass model \cite{[Gor07]} by the Bruxelles-Montr{\'e}al collaboration, data on fission barriers
were utilized to optimize the EDF parameters by adding phenomenological
collective corrections, including a rotational one. 
One may, therefore,  conclude that no EDF has ever been
systematically optimized at the deformed HFB level (and without phenomenological corrections added) by explicitly considering constraints on states at large  deformations.
 
In a previous study \cite{[Kor10]},  we applied modern optimization and statistical methods, 
together with high-performance computing, to carry out EDF optimization at the deformed HFB level, namely, the approximation level 
where the functional is later applied. The resulting EDF parameterization {\UNEDFZERO} yields good agreement
with experimental masses, radii, and deformations. The present work
represents an extension of \cite{[Kor10]} to the 
problem of fission. In particular, it builds on the results reported 
in Ref.~\cite{[Nik10]}, which concluded  that the data on strongly 
deformed nuclear states should be considered in the optimization protocol
to constrain the surface properties of the functional.

Here we  propose the new EDF Skyrme parameterization, {\UNEDFONE}, which is obtained by adding to the list of fit observables the experimental 
excitation energies of fission isomers in the actinides. 
To ensure that  the functional can be used  in fission and fusion studies,  
we have removed the center-of-mass (c.o.m.) correction in the spirit of the DFT. 
As in the case of  {\UNEDFZERO}, a sensitivity analysis has  been performed at the solution 
in order to identify possible correlations between model parameters and assess the impact of the new class of fit observables on the resulting  parameterization.

This paper is organized as follows. In Sec. \ref{Sec-theory} we briefly review 
the theoretical framework, establish the notation, and justify the removal of the c.o.m.\ correction. Section \ref{Sec-optimization}
defines the  set of fit observables, discusses numerical precision and implementation, and
presents the new  {\UNEDFONE} parameter set together with
the results of the sensitivity analysis.
To assess the impact of fission-isomer data, we compare {\UNEDFONE} with {\UNEDFZERO} in Sec.~\ref{Sec-parameters}.
In Sec. \ref{Sec-results} we study the performance of {\UNEDFONE} with respect to global nuclear observables, spectroscopic properties,  fission, and neutron droplets. Section~\ref{Sec-conclusions} contains the main conclusions and lays out future work.


\section{Theoretical Framework}
\label{Sec-theory}


\subsection{Time-Even Skyrme Energy Density Functional}
\label{Subsec-skyrme}

In the nuclear DFT, the total binding energy $E$ of the nucleus is a functional of the one-body density 
$\rho$ and pairing $\tilde{\rho}$ matrices. In its quasi-local 
approximation, it can be written as a  3D spatial integral:
\begin{eqnarray}
E[\rho,\tilde{\rho}] &=&  \int d^{3}\gras{r} ~ \mathcal{H}(\gras{r}) \nonumber \\
& = & \int d^{3}\gras{r} ~ \left[ \mathcal{E}^{\rm Kin}(\gras{r}) + \chi_{0}(\gras{r}) +
\chi_{1}(\gras{r}) 
\right. \nonumber \\ & & \left.
+ \tilde{\chi}(\gras{r}) + \mathcal{E}^{\rm Coul}_{\rm Dir}(\gras{r}) 
+ \mathcal{E}^{\rm Coul}_{\rm Exc}(\gras{r}) \right] \, ,
\label{EDFT}
\end{eqnarray}
where $\mathcal{H}(\gras{r})$ is the energy 
density that is quasi-local (it usually depends on derivatives with respect 
to the local densities), time-even, scalar, isoscalar, and real. It is usually broken down 
into the kinetic energy ($\mathcal{E}^{\rm Kin}(\gras{r})$) and nuclear  potential (for both the particle-hole and 
particle-particle channels, $\chi_{0,1}(\gras{r})$ and $\tilde{\chi}(\gras{r})$, respectively) 
and  Coulomb terms ($\mathcal{E}^{\rm Coul}_{\rm Dir}(\gras{r})$ and $\mathcal{E}^{\rm Coul}_{\rm Exc}(\gras{r})$). For Skyrme functionals, the 
particle-hole energy density $\chi_{0}(\gras{r})+\chi_{1}(\gras{r})$ splits into  
$\chi_{0}(\gras{r})$, depending only on isoscalar densities, and 
$\chi_{1}(\gras{r})$, depending  on isovector densities (and the isoscalar particle density through the density dependence of the coupling constant $C_1^{\rho\rho}$; see below) \cite{[Ben03],[Per04],[Roh10]}. Each term takes 
the generic form
\begin{eqnarray}
\chi_t(\gras{r}) & = & C_t^{\rho\rho} \rho_t^2 
  + C_t^{\rho\tau} \rho_t\tau_t +  C_t^{J^2} \bm{J}_t^2
\nonumber \\  &    &
  + C_t^{\rho\Delta\rho} \rho_t\Delta\rho_t \
  + C_t^{\rho \nabla J}  \rho_t\bm{\nabla}\cdot\bm{J}_t \, ,
\label{eq:UED}
\end{eqnarray}
where  $\rho_{t}$, $\tau_{t}$, and $\bm{J}_{t}$ ($t = 0,1$) can all be expressed in terms 
of full density matrix $\rho_{t}(\gras{r}\sigma,\gras{r'}\sigma')$; see Ref. \cite{[Ben03]} for details. (For brevity, we have omitted 
the explicit dependence of the densities on the coordinate $\gras{r}$.)
The $\bm{J}_{t}$ density is the vector part of the spin-current density tensor $J_{\mu\nu}$. (As in our previous work \cite{[Kor10]}, non-vector components of 
$J_{\mu\nu}$ were disregarded.)
The coupling constants are real numbers, except for 
$C_t^{\rho\rho}$, which is taken to be density-dependent:
\begin{equation}
C_t^{\rho\rho} =  C_{t0}^{\rho\rho}  + C_{t{\rm D}}^{\rho\rho}~ \rho_0^\gamma. 
\label{eq:crramp}
\end{equation}
All volume coupling constants ($C_{t}^{\rho\rho}$ and $C_{t}^{\rho\tau}$) can be 
related to the constants characterizing the infinite nuclear matter  \cite{[Kor10]}, and this relation was used 
during the optimization in order to define the range of parameter changes.

The Coulomb contribution  is treated as usual by assuming 
a point proton charge. The exchange term was computed at the Slater approximation:
\begin{equation}
\mathcal{E}^{\rm Coul}_{\rm Exc}(\gras{r}) = - \frac{3}{4} e^{2} 
\left(\frac{3}{\pi}\right)^{1/3}\rho_{\rm p}^{4/3} \, .
\end{equation}

For the pairing energy density $\tilde{\chi}(\gras{r})$, we use the mixed 
pairing description of \cite{[Dob02]} with
\begin{equation}
\tilde{\chi}(\gras{r}) = \sum_{q=n,p} \frac{V^{q}_{0}}{2}
\left[ 1 - \frac{1}{2}\frac{\rho_{0}(\gras{r})}{\rho_{\rm c}} \right]\tilde\rho^2(\gras{r}) \, ,
\label{eq:vpair}
\end{equation}
where  $\tilde\rho$ is the local pairing density. The value $\rho_{c}$=0.16\,
fm$^{-3}$ is used throughout this paper. We allow for different pairing 
strengths for protons ($V^{p}_0$) and neutrons ($V^{n}_0$) \cite{[Ber09a]}. A cut-off of 
$E_{\rm cut}=60\,{\rm MeV}$ was used to truncate the quasi-particle space 
\cite{[Dob84]}. To prevent the collapse of pairing, we used the Lipkin-Nogami 
procedure according to \cite{[Sto03]}.


\subsection{Treatment of the Center of Mass}
\label{Subsec-com}

The success of the self-consistent mean-field theory is, to a great extent, due 
to the concept of symmetry breaking. A classic example is the breaking of the 
translational invariance by the  mean field that is localized in space. The associated c.o.m.\ correction to the binding energy \cite{[Boh75],[RS80w],[Ben03]}, $-\langle\gras{\hat{P}}_{\rm c.m.}^2\rangle/(2mA)$, is usually added to the DFT binding energy in (\ref{EDFT}).
This correction contributes typically a few MeV to the total 
energy. Moreover, it was shown that adopting 
approximations to this correction during the optimization of the functional 
could lead to significantly different surface properties \cite{[Ben00],[Ben03],[Dob09g]}. 

Since the c.o.m.\ correction is not additive in particle number, it causes serious conceptual 
problems when dealing with fission or  heavy-ion fusion, that is, 
when one considers the  split of the nucleus into several fragments, or formation of the compound nucleus through a merger of two lighter ions.
In fission studies, it was shown that the 
contribution of the c.o.m.\  correction between the two prefragments 
could amount to several MeV near scission \cite{[Ber80],[Ska07],[Ska08]}. 
Moreover, properly computing this relative contribution is difficult, as it 
reflects the degree of entanglement between  prefragments \cite{[Ska08],
[You11]}. Time-dependent Hartree-Fock  calculations of low-energy heavy-ion reactions are even more 
problematic, as there is currently no solution to the discontinuity of the 
c.o.m.\ correction between the  target+projectile system and the compound 
nucleus \cite{[Uma06],[Uma09]}: such calculations usually neglect the c.o.m.\ 
term altogether, even though EDFs employed in such calculations have been usually fitted with the c.o.m. correction included. 
There are, however, some exceptions, see,  e.g.,  Ref.~\cite{[Kim97]}.
Note that the same problem occurs with the so-called rotational correction arising from the  
breaking of rotational invariance by deformed mean fields \cite{[Ber80],[Erl10]}.

Another undesired property of  the c.o.m.\ correction is that it slightly breaks the variational nature of HFB when adding or subtracting a particle \cite{[Zal08]}, i.e., it violates the Koopmans theorem. 
The resulting s.p. energy shifts are quite significant and they are of the order of the mass polarization effect  related to the fact that when adding or subtracting a particle to a closed 
spherical core, the resulting nucleus becomes deformed. If a spherical symmetry is imposed, 
the mass polarization effect is  a self-consistent rearrangement of all nucleons, when an odd particle
is introduced to the system. This corresponding energy  shift is $E_{\rm Kin}/A^{2}$, i.e.,    about 0.4~MeV in $^{40}$Ca \cite{[Zal08]}.

As discussed in Sec.~\ref{Sec-introduction}, the EDF is supposed to  capture all the physics of interest at the HFB level. In other words, the functional is to be built from 
the full single-reference density matrix.  While the HFB vacuum breaks the translational symmetry, the associated correction term should be absorbed in the density dependence. In particular, it should be 
possible to express the c.o.m.\ term  not as an explicit function of $A$, as currently being done, but through a density functional. Until a simple prescription is proposed, however, 
it is more consistent 
to simply drop all corrections that originate from a Hamiltonian view of the 
problem, including  the c.o.m.\ term. And, since our focus in on fission,  
this is precisely what we have done in this work.


\section{Optimization and Sensitivity Analysis}
\label{Sec-optimization}


\subsection{Experimental Dataset}
\label{Subsec-exp}

Since the focus of this study is the construction of an EDF optimized 
for fission, our experimental dataset has been expanded by including the excitation 
energies (bandheads) of four fission isomers (FIs) listed in Table \ref{table:newdata}. The ground-state (g.s.) binding energies of  $^{238}$U, $^{240}$Pu, and $^{242}$Cm, not 
included in the previous {\UNEDFZERO} dataset \cite{[Kor10]}, were added  for consistency (the g.s.\ energy of  $^{236}$U was already there). 
\begin{table}[ht]
\begin{center}
\caption{Experimental excitation energies of fission isomers \cite{[Sin02]}
(in MeV) considered in  the {\UNEDFONE}  dataset. 
}
\begin{ruledtabular}
\begin{tabular}{ccc}
Z & N & $E$ \\
92 & 144 &  2.750  \\
92 & 146 &  2.557  \\
94 & 146 &  2.800 \\
96 & 146 &  1.900  \\
\end{tabular}
\end{ruledtabular}
\label{table:newdata}
\end{center}
\end{table}
Consequently, compared with {\UNEDFZERO},  the {\UNEDFONE} dataset
contains seven new data points: three additional g.s.\ 
masses of deformed nuclei, and four excitation energies of FIs. 
For the FIs, we used the weight $w_i=0.5$\,MeV in the $\chi^2$ objective function
\begin{equation}
\chi^2(\xb)= \sum \limits_{i} \left(\frac{ s_{i}(\xb)-d_{i} } { w_i } \right)^2.
\label{eq:chi2}
\end{equation}
The $\chi^2$ weights for binding energies, proton rms radii, and odd-even 
mass (OEM) staggering  are the same as in  Ref.~\cite{[Kor10]}.

Two assumptions made in \cite{[Kor10]} were also adopted 
here: (i) since the isovector effective mass cannot be reliably 
constrained by the current data, it was set to $1/M_{v}^{*}=1.249$ as in 
{\UNEDFZERO} and the SLy4 parameterization \cite{[Cha98]}, which was the initial starting point in our optimization; 
and (ii) since tensor terms 
are mostly sensitive to the single-particle (s.p.) shell structure, 
which is not directly constrained by  the {\UNEDFONE}  dataset,
the tensor coupling constants 
$C_{0}^{J^2}$ and $C_{1}^{J^2}$ were set to zero. 
In summary, compared with {\UNEDFZERO} \cite{[Kor10]},  the optimization of {\UNEDFONE} is
 characterized by the following:
\begin{itemize}
\item The same 12 EDF parameters to be optimized, namely, $\rhoc$, $\enm/A$, $\knm$, $\asym$, $\lsym$, $M_s^*$,
$C_{0}^{\rho\Delta\rho}$, $C_{1}^{\rho\Delta\rho}$, $V_0^n$, $V_0^p$, $C_{0}^{\rho\nabla J}$, and  $C_{1}^{\rho\nabla J}$;
\item 7 additional data points: 3 new masses and 4 FI energies with the weights $w=0.5$\,MeV;
\item Neglect of the c.o.m.\ correction term.
\end{itemize}


\subsection{Numerical Precision and Implementation}
\label{Subsec-precision}

All HFB calculations were run with the code {\HFBTHO} \cite{[Sto05]}. The code 
expands the HFB solutions on the axially symmetric, deformed harmonic oscillator 
(HO) basis. In the optimization of {\UNEDFZERO}, we used a spherical basis with 20 HO shells, which was found to give a good compromise between the numerical 
precision and computational performance. The current optimization includes 
states with much larger deformation than in the ground state, and the dependence 
of the energies with respect to the set of basis states is more significant. 

\begin{figure}[ht]
\center
  \includegraphics[width=\linewidth]{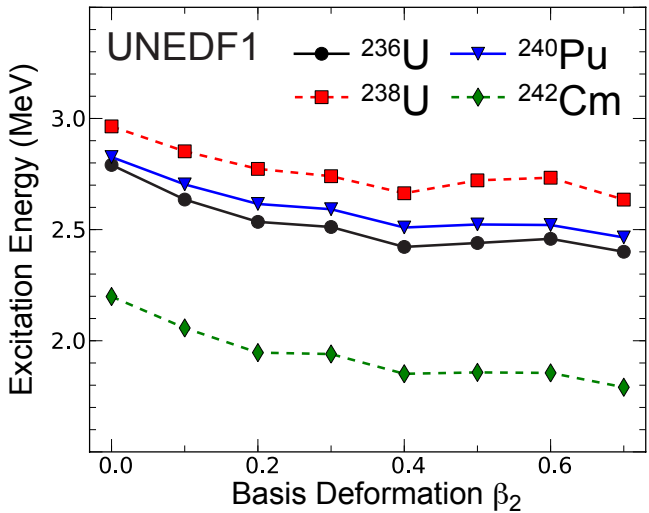}
  \caption{(Color online) Excitation energies of  fission 
           isomers considered  in the {\UNEDFONE} optimization as functions of 
           the HO basis deformation.}
  \label{fig:numerics}
\end{figure}

In the {\UNEDFONE} optimization, all quantities but the four fission isomers were computed with 
the  spherical HO basis of $N_{sh}=20$   shells, which includes  $N=1771$ basis 
states. For the fission isomers, we adopted a stretched HO basis  with deformation $\beta = 0.4$. 
The basis contains  up to $N_{sh}=50$ oscillator shells with an upper limit of 
$N=1771$ basis states with lowest HO s.p.\ energies. 
The  oscillator frequency $\omega_{0}^{3} = \omega_{\perp}^{2} 
\omega_{\parallel}$ was set at $\hbar\omega_{0} = 1.2 \times 41/A^{1/3}$\,MeV.  
As seen in Fig. \ref{fig:numerics}, at this selection of the HO basis, the dependence of FI energies on the basis deformation  remains fairly constant around $\beta = 0.4$. Moreover, the range of variations is significantly less than the corresponding  $\chi^2$ weight, $w_i=0.5$\,MeV.

Optimization calculations were performed on Argonne National Laboratory's
Fusion cluster, managed by Argonne's Laboratory Computing Resource Center (LCRC).
Fusion consists of 320 computing nodes, each  with dual quad-core Pentium Xeon processors. By using Intel's Math Kernel Library and  the Intel Fortran compiler (ifort), we were
able to run {\HFBTHO} in almost half the time when compared with the prebuilt
reference BLAS library implementation and GNU's gfortran compiler.  We were 
also able to dramatically reduce the wall-clock time of an {\HFBTHO} computation by 
using OpenMP at the node level to parallelize key computational bottlenecks. 
These bottlenecks 
involved iteratively computing the eigenvalues and eigenvectors of the 
($\Omega$, $\pi$) blocks of the HFB matrix, as well as the density calculations 
reflecting the same block pattern. OpenMP allowed us to dynamically assign 
processors to blocks of data for parallel processing, which further reduced the 
wall-clock time by a factor of 6 when running on an eight-core node. 

The parameter estimation computations presented in this paper ran 218 total simulations of {\HFBTHO} 
for each nucleus in the dataset, using 80 compute nodes (640 cores) for 5.67 
hours.
As highlighted in \cite{[Kor10]}, using the {\algo} algorithm (Practical Optimization Using No Derivatives (for Squares)) on the type of fitting problem considered here requires more than 10 times fewer {\HFBTHO} runs over a more traditional, derivative-free Nelder-Mead optimization method 
\cite{[Nel65]}.  Hence, without the algorithmic and computational advancements detailed above, a similar optimization could have previously consumed a month of computations using 80 cores of the Fusion cluster.

We emphasize that, strictly speaking, both the 
 {\UNEDFZERO} and the {\UNEDFONE} parameterizations obtained in this work should always be 
used in their original environment. 
In particular, the pairing EDF should  be that  of Eq.~(\ref{eq:vpair}) used with the original pairing space cut off; pairing calculations must  be complemented 
by the Lipkin-Nogami prescription; and the proton and neutron pairing strengths 
must not vary from the values determined by our optimization. In short, contrary to usual practice,  there is no flexibility in the treatment of the pairing channel.
 

\subsection{Result of the Optimization: {\sc\bfseries UNEDF1} Parameter Set}
\label{Subsec-optimization}

The starting point for our  {\algo} optimization was the previously
obtained {\UNEDFZERO} parameterization. After 177 simulations, the algorithm 
reached the new optimal result. The resulting parameter set is listed 
in Table \ref{table:optimum}. The first six parameters were restricted to
lie within finite bounds, also listed in Table \ref{table:optimum},
that were not allowed to be violated during the optimization 
procedure.
As can be seen, parameters 
$ \enm/A $ and $ \knm $ are on the boundary value. In the case of {\UNEDFZERO}, 
we recall that $\knm$ and $ 1/M_{s}^{*} $ also ended up at their respective boundaries. 
The saturation density $\rho_c$ is given with more digits than the other parameters. Such extra precision is needed when computing volume coupling constants \cite{[Kor10]}.  
 
\begin{table}[ht]
\begin{center}
\caption{Optimized parameter set {\UNEDFONE}. Listed are bounds used in the 
         optimization, final optimized parameter values, standard deviations, and 
         95\% confidence intervals.}
\begin{ruledtabular}
\begin{tabular}{lcrrr}
$ \xb $                     & Bounds  & $ \hat{\xb}^{\rm (fin.)} $ &
$\sigma$ & 95\% CI \\
\hline
$ \rhoc $                  & [0.15,0.17]        &  0.15871  &
0.00042 & [   0.158,   0.159] \\
$ \enm/A $                 & [-16.2,-15.8]        &   -15.800 &     --
&          --         \\
$ \knm $                   & [220, 260]         &   220.000 &     --
&          --         \\
$ \asym $                  & [28, 36]        &    28.987 &  0.604
& [  28.152,  29.822] \\
$ \lsym $                  & [40, 100]        &    40.005 & 13.136
& [  21.841,  58.168] \\
$ 1/M_{s}^{*} $            & [0.9, 1.5]        &     0.992 &  0.123
& [   0.823,   1.162] \\
$ C_{0}^{\rho\Delta\rho} $ & [$-\infty, +\infty$] &   -45.135 &  5.361
& [ -52.548, -37.722] \\
$ C_{1}^{\rho\Delta\rho} $ & [$-\infty, +\infty$] &  -145.382 & 52.169
& [-217.515, -73.250] \\
$ V_0^n $                  & [$-\infty, +\infty$] &  -186.065 & 18.516
& [-211.666,-160.464] \\
$ V_0^p $                  & [$-\infty, +\infty$] &  -206.580 & 13.049
& [-224.622,-188.538] \\
$ C_{0}^{\rho\nabla J}$    & [$-\infty, +\infty$] &   -74.026 &  5.048
& [ -81.006, -67.046] \\
$ C_{1}^{\rho\nabla J}$    & [$-\infty, +\infty$] &   -35.658 & 23.147
& [ -67.663,  -3.654] \\
\end{tabular}
\end{ruledtabular}
\label{table:optimum}
\end{center}
\end{table}

We first note that the same minimum was obtained by starting either from 
the {\UNEDFZERO} solution or from the {\UNEDFONE}ex parameterization discussed below: this gives us 
confidence that the parameter set listed in Table \ref{table:optimum} is 
sufficiently robust. We can then observe that most of the parameter values of {\UNEDFONE} 
are fairly close to those of {\UNEDFZERO} \cite{[Kor10]}. There are, nevertheless, 
a couple of notable exceptions. First, the magnitude of 
$C_{1}^{\rho\Delta\rho}$ is now much larger. This is potentially dangerous, 
as it might trigger scalar-isovector instabilities in the functional that 
could appear in neutron-rich nuclei \cite{[Les06],[Kor10b]}. (Our mass-table calculations with {\UNEDFONE}
do not show indications of instability in even-even nuclei.)
Second, 
$C_{1}^{\rho\nabla J}$ has drifted considerably from its initial value, even 
changing sign. These two coupling constants control the 
isovector surface properties of the nucleus; hence, only proper 
constraints on the shell structure like, for example, spin-orbit splitting in 
neutron-rich nuclei will allow these terms to be pinned down. For the moment, 
both coupling constants are relatively unconstrained, as evidenced also by their relatively 
large $\sigma$-value shown in Table \ref{table:optimum}.


\subsection{Sensitivity Analysis}
\label{Subsec-sensitivity}


\subsubsection{Correlation Matrix of {\UNEDFONE}}
\label{Subsubsec-sensUNEDF1}

\begin{table*}[!]
\begin{center}
\caption{Correlation matrix for {\UNEDFONE} parameter set. Correlations greater 
         than 0.8 (in absolute value) are in boldface.}
\begin{ruledtabular}
\begin{tabular}{lrrrrrrrrrr}
$ \rhoc $                  &  1.00  \\
$ \asym $                  & -0.35 &  1.00 \\
$ \lsym $                  & -0.14 &  0.71 &  1.00 \\
$ 1/M_{s}^{*} $            &  0.32 &  0.23 &  0.36 & 1.00 \\
$ C_{0}^{\rho\Delta\rho} $ & -0.25 & -0.25 & -0.35 & {\bf -0.99} & 1.00 \\
$ C_{1}^{\rho\Delta\rho} $ & -0.06 & -0.15 & -0.77 &      -0.22  &       0.19  & 1.00 \\
$ V_0^n $                  & -0.32 & -0.22 & -0.36 & {\bf -0.99} & {\bf  0.98} & 0.22 & 1.00 \\
$ V_0^p $                  & -0.33 & -0.18 & -0.29 & {\bf -0.97} & {\bf  0.97} & 0.15 & {\bf 0.96} & 1.00 \\
$ C_{0}^{\rho\nabla J}$    & -0.14 & -0.20 & -0.32 & {\bf -0.86} & {\bf  0.91} & 0.22 & {\bf 0.85} & {\bf 0.84} &  1.00 \\
$ C_{1}^{\rho\nabla J}$    & 0.05 & -0.17 & -0.13 &      -0.10  &       0.07  & 0.21 &      0.10  &      0.07  & -0.03 & 1.00 \\
\hline
 & $ \rhoc $ & $ \asym $ & $\lsym$ & $ 1/M_{s}^{*} $ & $ C_{0}^{\rho\Delta\rho} $ & $ C_{1}^{\rho\Delta\rho} $ &
   $ V_0^n $ & $ V_0^p $ & $ C_{0}^{\rho\nabla J}$ & $ C_{1}^{\rho\nabla J}$ \\
\end{tabular}
\end{ruledtabular}
\label{table:correlation}
\end{center}
\end{table*}

We have performed a sensitivity analysis at the solution of the 
optimization. All residual derivatives were estimated by using the optimal 
finite-difference procedure detailed in Ref.~\cite{[Mor11]}. 
Since some of the parameters ran at their bounds during the 
optimization, the sensitivity analysis was carried out in a subspace that 
does not contain these parameters. The same strategy was also used in the 
previous sensitivity analysis of the {\UNEDFZERO} parameterization; we refer to 
\cite{[Kor10]} for a detailed discussion of the available options in the case 
of constrained optimization. In Table \ref{table:optimum} we list the standard 
deviation, $\sigma$, and 95\% confidence interval (CI) for each parameter at the solution. 
As discussed in Sec.~\ref{Subsec-optimization}, the standard deviations of most of the parameters is relatively small.

\begin{figure}[ht]
\center
  \includegraphics[width=1.0\linewidth]{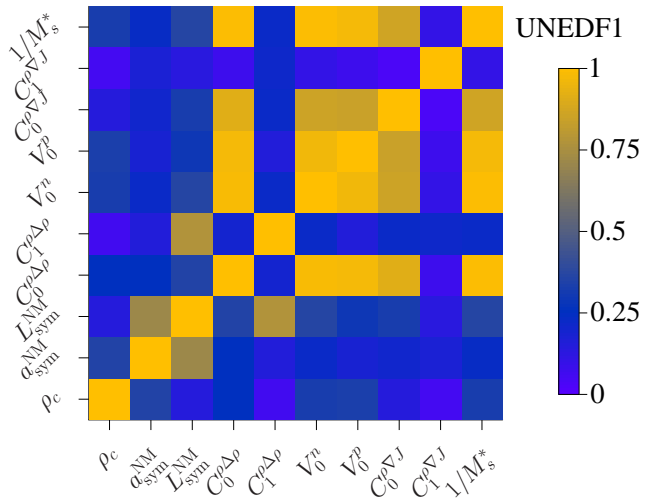}
  \caption{(Color online) Absolute values of the correlation matrix of Table~\ref{table:correlation}
   presented in a color-coded format.}
  \label{fig:correlation}
\end{figure}

The correlation matrix for the {\UNEDFONE} parameter set is presented in 
Table \ref{table:correlation}. It was calculated as in \cite{[Kor10]} and,  
similarly, corresponds to the 10-dimensional subspace of the 
parameters that are not at their boundary value. Generally, most of the 
parameters are only slightly correlated to each other, with a few notable 
exceptions (correlations below 0.8 are not very significant 
from a statistical viewpoint). The strong correlation between $ 1/M_{s}^{*}$ 
and both $V_0^n$ and $V_0^p$ had already been noticed in the {\UNEDFNB} 
parameter set of Ref.~\cite{[Kor10]} and reflects the interplay between the level 
density at  the Fermi-surface and the size of pairing correlations. 
Similarly, both pairing strengths are strongly correlated 
with $ C_{0}^{\rho\Delta\rho}$, which can also be related to surface properties of the 
functional. Interestingly, both pairing strengths are now strongly correlated 
with one another, which was not the case with  {\UNEDFZERO}.
The same correlation matrix of Table~\ref{table:correlation} is shown
graphically in Fig.~\ref{fig:correlation}.

\begin{figure}[ht]
\center
  \includegraphics[width=\linewidth]{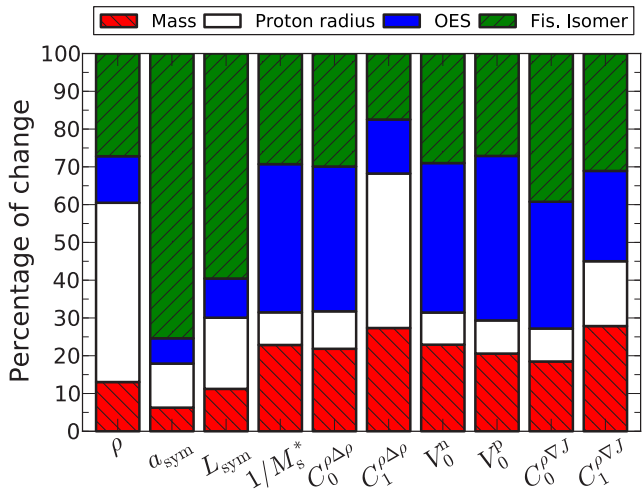}
  \caption{(Color online) Sensitivity of {\UNEDFONE} to different types of data    
           entering the $\chi^{2}$ function. }
  \label{fig:partsum}
\end{figure}

Next, we  study the overall impact of each data type in our $\chi^{2}$ 
function on the obtained parameter set. As in \cite{[Kor10]} we  
calculate the partial sums of the sensitivity matrix for each data type. 
Let us recall that the sensitivity  matrix $S$ is defined as
\begin{equation}
S(\xb)=\left[ J(\xb)J^{\rm T}(\xb) \right]^{-1}J(\xb) \, ,
\end{equation}
where $J(\xb)$ is the Jacobian matrix. The results are illustrated in Fig.~\ref{fig:partsum}, where we have summed absolute values  of each data type for each parameter. The total strengths for each parameter were then normalized to 100\%. Note that the  fission isomer excitation energies represent less than 4\% of the total number of data points but account for typically 30\% of the variation of the parameter set. In the case of the symmetry energy coefficient, this percentage is even 75\% (see Sec.~\ref{Sec-parameters} for more discussion). Compared with {\UNEDFZERO}, we find that the overall dependence on the proton radii has significantly decreased, except for $\rho$ and $C_{1}^{\rho\Delta\rho}$, and that the dependence on the OES has actually increased. This kind of analysis, however, does not address the importance of a particular data point to the obtained optimal solution. 

\begin{figure}[ht]
\center
  \includegraphics[width=1.0\linewidth]{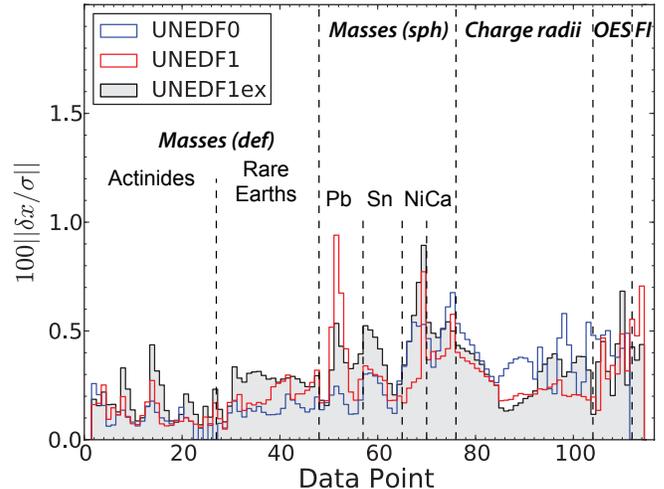}
  \caption{(Color online) Overall change in $\xb$ for the {\UNEDFONE} when the datum $d_{i}$ is changed by an amount of $0.1w_{i}$ one by one. The four rightmost data points marked FI correspond to excitation energies of fission isomers. The results for {\UNEDFZERO} and {\UNEDFONE}ex of Sec.~\ref{Subsubsec-sensUNEDF1ex} are also shown.}
  \label{fig:sensitivity_UNEDF1}
\end{figure}

A complementary way to study the impact of an individual datum on the obtained 
parameter set is therefore presented in Fig. \ref{fig:sensitivity_UNEDF1}. Here, we have plotted the amount of variation
\begin{equation}
\vert\vert {\delta\xb}/{\sigma} \vert\vert 
= \sqrt{\sum_{k}\left(\frac{\delta x_{k}}{\sigma}\right)^{2}}
\end{equation}
for the optimal parameter set when data points $d_{i}$ are changed by an amount of $0.1w_{i}$ one by one. As can be seen, the variations are small overall, assuring us that the dataset was chosen correctly. The masses of the double magic nuclei $^{208}$Pb and $^{58}$Ni seem to have the biggest relative impact on the optimal parameter set. One can also see that the sensitivity of the parameters on the new FI data is larger than the average datum point. By contrast, the dependence of the parameterization on the masses of deformed actinide and rare earth nuclei is weaker.


\subsubsection{Discussion of the Coulomb Exchange Term}
\label{Subsubsec-sensUNEDF1ex}

It has been argued in \cite{[Gor08]} that removing the Coulomb exchange term 
from the functional could  improve the overall fit on nuclear 
binding energies. This procedure had been motivated by the earlier works  
Refs.~\cite{[Bro98],[Fay98]}, in which similar ameliorations, albeit on a smaller 
dataset, were observed. The origin of these ad hoc manipulations was the observation 
that many-body effects induced by the long-range Coulomb force among protons 
manifest themselves in the form of a (positive) correlation energy, which, to 
some extent, can cancel out the (negative) exchange term \cite{[Bul96],[Bul99]}. 
Since such an exchange-correlation effect is absent from the standard Skyrme functional, one 
could feel justified to simulate it by effectively screening the Coulomb exchange term 
with an empirical factor $0\leq \alpha_{\text{ex}} \leq 1$. The special cases 
$\alpha_{\text{ex}}=0$ and $\alpha_{\text{ex}}=1$ give, respectively, the 
case without and with full Coulomb exchange.

\begin{table}[ht]
\begin{center}
\caption{Optimized parameter set {\UNEDFONE}ex. Listed are bounds used in the 
         optimization, final optimized parameters, standard deviations, and 
         95\% confidence intervals.}
\begin{ruledtabular}
\begin{tabular}{lcrrr}
$ \xb $                     & Bounds  & $ \hat{\xb}^{\rm (fin.)} $ & $\sigma$ & 95\% CI \\
\hline
$ \rhoc $                  & [0.15, 0.17]        &  0.15837  & 0.00049 & [   0.158,   0.159] \\
$ \enm/A $                 & [-16.2, -15.8]        &   -15.800 &     --  &          --         \\
$ \knm $                   & [220, 260]         &   220.000 &     --  &          --         \\
$ \asym $                  & [28, 36]        &    28.384 &  0.711  & [  27.417,  29.351] \\
$ \lsym $                  & [40, 100]        &    40.000 &     --  &          --         \\
$ 1/M_{s}^{*} $            & [0.9, 1.5]        &     1.002 &  0.123  & [   0.835,   1.169] \\
$ C_{0}^{\rho\Delta\rho} $ & [$-\infty, +\infty$] &   -44.602 &  5.349  & [ -51.872, -37.331] \\
$ C_{1}^{\rho\Delta\rho} $ & [$-\infty, +\infty$] &  -180.956 & 47.890  & [-246.050,-115.863] \\
$ V_0^n $                  & [$-\infty, +\infty$] &  -187.469 & 18.525  & [-212.649,-162.288] \\
$ V_0^p $                  & [$-\infty, +\infty$] &  -207.209 & 13.106  & [-225.024,-189.395] \\
$ C_{0}^{\rho\nabla J}$    & [$-\infty, +\infty$] &   -74.339 &  5.187  & [ -81.389, -67.289] \\
$ C_{1}^{\rho\nabla J}$    & [$-\infty, +\infty$] &   -38.837 & 23.435  & [ -70.690,  -6.984] \\
$ \alpha_{\text{ex}}$       & [    0,    1]        &     0.813 &  0.154  & [   0.604,   1.023] \\
\end{tabular}
\end{ruledtabular}
\label{table:optimumUNEDF1ex}
\end{center}
\end{table}

The result of the optimization of the functional with this additional 
parameter $\alpha_{\text{ex}}$ is given in Table \ref{table:optimumUNEDF1ex}. 
Our objective function is slightly decreased from 51.058 to 49.341 when this 
term is present. Overall, both parameterizations, with and without the 
Coulomb exchange screening term, are very similar. However, one can see 
that the 95\% CI is relatively large for the screening parameter, the value of which is also close to 1 (full Coulomb exchange). We recall 
that this confidence interval is extracted from the correlation matrix 
computed in the 10-D space of ``inactive'' parameters, namely, the space of 
the 10 parameters that are not at their bound and thus actively constrained. 
If one computes the Jacobian matrix in the original 13-D space of all parameters 
with a tangent plane approximation to account for the three active parameters 
$\enm$, $\knm$, and $\lsym$, we find that the 95\% CI for the screening 
parameter becomes [-1.663,3.290]. This implies that $\alpha_{\text{ex}}$ 
is basically not constrained with the current dataset.

\begin{figure}[ht]
\center
  \includegraphics[width=\linewidth]{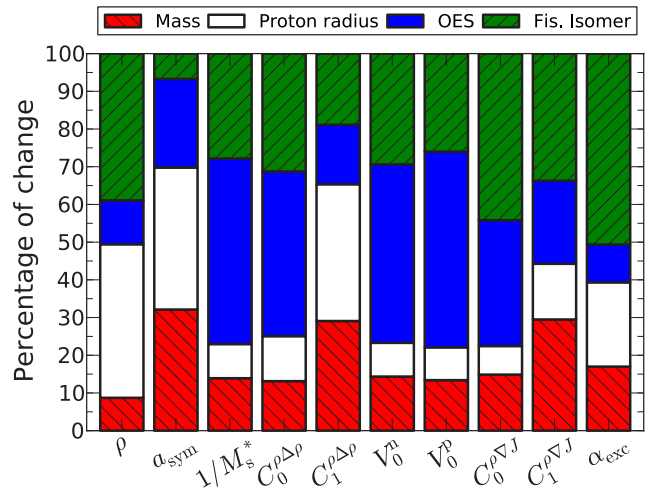}
  \caption{(Color online) Similar as in Fig.~\ref{fig:partsum} but for the {\UNEDFONE}ex parameter set.}
  \label{fig:sensitivity_UNEDF1ex}
\end{figure}

The dependence of every parameter on the four types of data included in the 
dataset (masses, charge  radii, OES, and FI data) 
reveal an interesting consequence of the screening of the Coulomb exchange 
term. Figure~\ref{fig:sensitivity_UNEDF1ex} shows the analogue of Fig. 
\ref{fig:partsum} when the screening parameter $\alpha_{\text{ex}}$ 
is included. Note the striking difference in the bar plot for the symmetry 
energy parameter $\asym$: fluctuations in this parameter under 
a variation of the excitation energy of the fission isomers are reduced 
to less than 10\%, compared with nearly 75\% when the full Coulomb 
exchange term is computed.

This behavior can be qualitatively understood by recalling a few simple 
facts about the bulk nuclear energy and deformation. A variation of the 
excitation energy of very deformed states such as  fission isomers essentially 
affects the bulk surface properties of the functional---especially if 
the coupling constants driving shell effects are somewhat constrained by 
the dataset. In the language of the leptodermous expansion of Sec.~\ref{Subsec-LDM}, this implies 
that both the  surface and surface-symmetry energy coefficients (which 
depend in a  nontrivial way on the coupling constants of the functional) 
should be impacted. On the other hand, we may assume that the isospin 
dependence of the binding energy (i.e., the total symmetry energy) is 
relatively well constrained by the several long isotopic 
sequences present in our dataset. We therefore see that the requirement of having the full symmetry 
energy constrained together with a relatively large variation of the 
surface terms should lead to a relatively large variation of the volume 
symmetry $\asym$, which is indeed observed in Fig. \ref{fig:sensitivity_UNEDF1}.

One can now understand the difference of behavior of $\asym$ under a change 
of data when the Coulomb screening term is present: according to \cite{[Bul96],
[Bul99]}, the many-body Coulomb correlation energy that is simulated by  
$\alpha_{\text{ex}}<1$  essentially represents a proton surface effect. 
Changes in bulk surface properties triggered by variations in the excitation 
energy of fission isomers can be entirely absorbed by a readjustment of 
$\alpha_{\text{ex}}$, especially since the latter is poorly constrained by 
the other data, rather than by $\asym$.
Lastly, we note that the Coulomb exchange term, which is  approximated  by the usual  local
Slater expression, may get worse at large
deformations \cite{[Leb11]}.

In summary, considering that (i) $\alpha_{\text{ex}}$ is poorly constrained by the data,  yet may affect significantly other parameters like $\asym$ and (ii) $\alpha_{\text{ex}}$  does not significantly improve the quality of the fit, we decided to retain the full Coulomb exchange term in the present {\UNEDFONE} parameterization.


\section{Characterization of {\UNEDFONE} parameterization}
\label{Sec-parameters}

In this section, we discuss general properties of the {\UNEDFONE} parameterization and compare it with {\UNEDFZERO}. 


\subsection{Energy Density in $\gras{(t,x)}$ Parameterization}
\label{Subsec-xt}

For practical applications, it is useful to express the coupling constants of {\UNEDFZERO} and {\UNEDFONE}  in the traditional  $(t,x)$-parameterization of the standard Skyrme force, see Appendix A of \cite{[Ben03]}. 
The results are given in Table~\ref{table:t_x}.
\begin{table}[ht]
\begin{center}
\caption{Parameters $(t,x)$ of {\UNEDFZERO} and {\UNEDFONE}.}
\begin{ruledtabular}
\begin{tabular}{lrrc}

Par. &   {\UNEDFZERO} &    {\UNEDFONE} & Units      \\\hline
$t_{0}$ & $-$1883.68781034 & $-$2078.32802326 & MeV$\cdot$fm$^{3}$ \\
$t_{1}$ &  277.50021224 &  239.40081204 & MeV$\cdot$fm$^{5}$ \\
$t_{2}$ &  608.43090559 & 1575.11954190 & MeV$\cdot$fm$^{5}$ \\
$t_{3}$ &13901.94834463 &14263.64624708 & MeV$\cdot$fm$^{3+3\gamma}$ \\
$x_{0}$ &    0.00974375 &    0.05375692 &     -      \\
$x_{1}$ &    $-$1.77784395 &    $-$5.07723238 &     -      \\
$x_{2}$ &    $-$1.67699035 &    $-$1.36650561 &     -      \\
$x_{3}$ &    $-$0.38079041 &    $-$0.16249117 &     -      \\
$b_{4}$ &  125.16100000 &   38.36807206 & MeV$\cdot$fm$^{5}$\\
$b_{4}'$ &    $-$91.2604000 &   71.31652223 & MeV$\cdot$fm$^{5}$\\
$\gamma$ &     0.32195599 &    0.27001801 &     -      \\
\end{tabular}
\end{ruledtabular}
\label{table:t_x}
\end{center}
\end{table}
As can be seen, in the $(t,x)$-parameterization the two functionals are quite different. 
This is to be expected as the relation between the 
$C$ and $(t,x)$ parameterizations is partially nonlinear~\cite{[Kor10a]}.

\subsection{Energy Density Parameters in Natural Units}
\label{Subsec-nu}
The EDF parameters can also be  expressed in terms of natural units 
\cite{[Kor10a]}. In Table \ref{table:natunit} we list the parameter set of  
{\UNEDFONE} in standard units and in natural units. Here we have used the same 
value for the scale  $\Lambda=687$\,MeV, characterizing the breakdown of the chiral effective theory, which has been found in 
Ref.~\cite{[Kor10a]}. From the numbers in Table \ref{table:natunit} one can 
see that most of the {\UNEDFONE} parameters are natural, with only 
two minor exceptions. First, because the effective mass 
$M_{s}^{*}$ in our optimum is close to unity, the $C_{0}^{\rho\tau}$ is abnormally  
small. Second, $C_{1}^{\rho\Delta\rho}$ seems be on the borderline of being 
unnaturally large. 
As Table \ref{table:optimum} indicates, however, the standard deviation for this  parameter is  rather large. It has to be noted, however, that
 there is nothing unusual about the magnitude of $C_{1}^{\rho\Delta\rho}$. Indeed, some examples of EDF
parameterizations with similar or larger values of $C_{1}^{\rho\Delta\rho}$ can be found in Fig.~2 of \cite{[Kor10a]}.
\begin{table}[ht]
\begin{center}
\caption{Coupling constants of {\UNEDFZERO} and {\UNEDFONE} in normal 
units and in natural units. Value $\Lambda=687\,{\rm MeV}$ was used.}
\begin{ruledtabular}
\begin{tabular}{lrrrr}
& \multicolumn{2}{c}{{\UNEDFZERO}} & \multicolumn{2}{c}{{\UNEDFONE}} \\
Coupling & Normal   & Natural & Normal   & Natural \\
Constant &  Units   &  Units  & Units    &  Units \\
\hline
$ C_{00}^{\rho\rho} $       & -706.38 & -0.795 & -779.37 & -0.878 \\
$ C_{10}^{\rho\rho} $       &  240.26 &  0.271 &  288.01 &  0.324 \\
$ C_{0{\rm D}}^{\rho\rho} $ &  868.87 &  0.901 &  891.48 &  0.937 \\
$ C_{1{\rm D}}^{\rho\rho} $ &  -69.77 & -0.072 & -201.37 & -0.212 \\
$ C_{0}^{\rho\tau} $        &  -12.92 & -0.176 &   -0.99 & -0.014 \\
$ C_{1}^{\rho\tau} $        &  -45.08 & -0.616 &  -33.52 & -0.458 \\
$ C_{0}^{\rho\Delta\rho} $  &  -55.26 & -0.755 &  -45.14 & -0.616 \\
$ C_{1}^{\rho\Delta\rho} $  &  -55.62 & -0.759 & -145.38 & -1.985 \\
$ C_{0}^{\rho\nabla J} $    &  -79.53 & -1.086 &  -74.03 & -1.011 \\
$ C_{1}^{\rho\nabla J} $    &   45.63 &  0.623 &  -35.66 & -0.487 \\
$ \gamma $                  &  0.3220 &        &  0.2700 &  \\
\end{tabular}
\end{ruledtabular}
\label{table:natunit}
\end{center}
\end{table}


\subsection{Leptodermous Expansion}
\label{Subsec-LDM}

To extract global properties of the energy functional and relate them to the familiar constants of the liquid 
drop model (LDM), one needs to carry out the leptodermous expansion. The general strategy behind the expansion of nuclear EDF has been discussed in Ref.~\cite{[Rei06]}, where one can find the relevant literature on this topic. The starting point is the LDM binding energy per nucleon expanded in the inverse radius ($\propto A^{-1/3}$) and neutron excess $I = (N-Z)/A$:
\begin{equation}
\begin{array}{rclclcl}
  {\cal E}(A,I)
  &=& \displaystyle
  a_{\rm vol} 
  &+& \displaystyle
  a_{\rm surf}A^{-1/3}
  &+& \displaystyle
  a_{\rm curv}A^{-2/3} \medskip\\
  &&&+&
  a_{\rm sym}{I^2}
  &+& \displaystyle
  a_{\rm ssym} A^{-1/3}{I^2} \medskip\\
  &&&&
  &+&  \displaystyle
  a_{\rm sym}^{(2)}I^4.
\end{array}
\label{eq:ldm}
\end{equation}
For any functional, our approach consists of combining nuclear matter (NM) 
calculations with Hartree-Fock (HF) calculations for a large set of 
spherical nuclei to extract by linear regression the various parameters 
of the expansion (\ref{eq:ldm}) according to the following procedure.

\begin{table}[ht]
\caption{\label{tab:nucmat}
Liquid drop coefficients of {\UNEDFZERO} and {\UNEDFONE}
(all in MeV).}
\begin{ruledtabular}
\begin{tabular}{l|ccc|ccc}
  & $a_{\rm vol}$ & $a_{\rm sym}$ & $a_{\rm sym}^{(2)}$ 
  & $a_{\rm surf}$ & $a_{\rm curv}$& $a_{\rm ssym}$ \\
\hline
{\UNEDFZERO} & -16.056 & 30.543 & 4.418 & 18.7 & 7.1 & -44 \\
{\UNEDFONE}  & -15.800 & 28.987 & 3.637 & 16.7 & 8.8 & -29 \\
\end{tabular}
\end{ruledtabular}
\end{table}

First, the bulk parameters $a_{\rm vol}$ and $a_{\rm sym}$ are 
directly obtained from NM calculations. Second, the smooth energy 
per nucleon $\bar{\cal E}(A,I)$ is extracted from the spherical HF calculations 
of $(A,I)$ nuclei by removing the shell correction~\cite{[Rei06]}. The isoscalar coefficients 
of the expansion (\ref{eq:ldm}) can then be deduced from the smooth energy by 
plotting
\begin{equation}
  \left[ \bar{\cal E}(A,0) - a_{\rm vol}\right]A^{1/3}
  \longrightarrow
  a_{\rm surf} + a_{\rm curv}A^{-1/3}
\label{eq:isoscalar}
\end{equation}
as a function of $A^{-1/3}$. The $a_{\rm surf}$ coefficient is obtained as the extrapolation 
of the curve to $A^{-1/3}\longrightarrow 0$. The curvature coefficient 
$a_{\rm curv}$ is then estimated from the slope of the line.

\begin{figure}[t]
\center
\includegraphics[width=\linewidth]{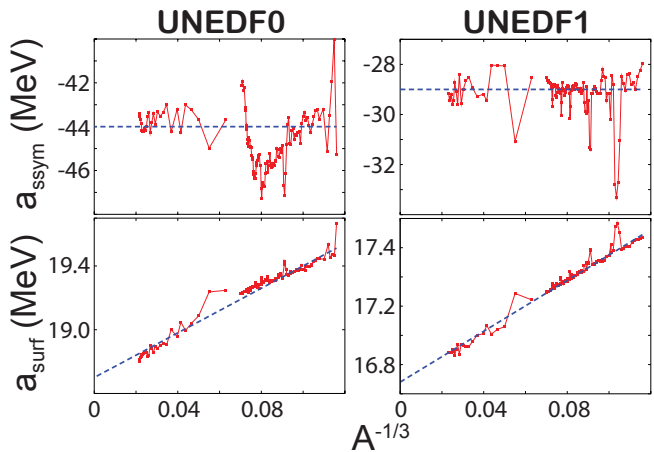}
\caption{(Color online) Surface-symmetry coefficient (\ref{eq:asymAverage1}) 
         (upper panels) and surface coefficient (lower panels) versus  $A^{-1/3}$ 
         for {\UNEDFZERO} (left) and {\UNEDFONE} (right). 
         }
\label{fig:asymm-surf} 
\end{figure}

The determination of isovector coefficients starts with the second-order symmetry 
coefficient $a_{\rm sym}^{(2)}$. It is easily estimated by systematic calculations 
in asymmetric NM. Defining
\begin{equation}
\begin{array}{rcl}
a_{\rm sym}^{\rm (eff)}(\infty,I) &= &\left[{\cal E}^{\rm NM}(\infty,I) - {\cal E}^{\rm NM}(\infty,0)\right]/{I^{2}} \\
& \longrightarrow &
a_{\rm sym} + a_{\rm sym}^{(2)}I^{2},
\label{eq:asymEffNM}
\end{array}
\end{equation}
one can extract the second-order symmetry coefficient from the slope  of 
$a_{\rm sym}^{\rm (eff)}(\infty,I)$ vs. $I^{2}$. Extracting the surface-symmetry 
coefficient is more involved. We first introduce the effective symmetry 
coefficient for a finite nucleus as
\begin{equation}
\begin{array}{rcl}
  a_{\rm sym}^{\rm (eff)}(A,I) & = & 
  \left[ \bar{\cal E}(A,I) - \bar{\cal E}(A,0) \right]/I^{2} \medskip\\
  & \longrightarrow & 
  a_{\rm sym} + a_{\rm ssym} A^{-1/3} + a_{\rm sym}^{(2)}I^2.
\label{eq:asymEff}
\end{array}
\end{equation}
In nuclear matter ($A\rightarrow +\infty$), the effective symmetry coefficient 
reduces to (\ref{eq:asymEffNM}). To avoid multidimensional regression analysis, 
we  introduce the reduced symmetry coefficient by subtracting the $I^2$-dependent 
part of the NM limit to $a_{\rm sym}^{\rm (eff)}$:
\begin{equation}
  a_{\rm sym}^{\rm (red)}(A,I) = 
  \left[ \bar{\cal E}(A,I) - \bar{\cal E}(A,0) \right]/I^{2} - a_{\rm sym}^{(2)}I^2.
\label{eq:asyRed}
\end{equation}
At the perfect LDM limit, the quantity $a_{\rm sym}^{\rm (red)}(A,I)$ 
should not depend on the neutron excess. At small isospins, however, numerical 
uncertainties in the shell-correction procedure are amplified by the $I^{2}$ 
denominator. In practice, it is more efficient to build an $I$-averaged reduced 
asymmetry coefficient,
\begin{equation}
  \overline{a_{\rm sym}^{\rm (red)}}
  = \frac{1}{b-a} \int_{a}^{b}dI\,a_{\rm sym}^{\rm (red)}(A,I),
\label{eq:asymAverage}
\end{equation}
where we choose $a = 0.1$ and $b = 0.2$~\cite{[Rei06]}. The surface-symmetry energy is then obtained from
\begin{equation}
  \left[\overline{a_{\rm sym}^{\rm (red)}} - a_{\rm sym}\right]A^{1/3} = a_{\rm ssym}.\label{eq:asymAverage1}
\end{equation}

Figure \ref{fig:asymm-surf} illustrates the numerical accuracy of the method 
of evaluation for the surface and surface-symmetry coefficients of the LDM. 
The dashed blue lines indicate the fitting lines from which the final values of 
$a_\mathrm{surf}$, $a_\mathrm{curv}$, and $a_\mathrm{ssym}$ are deduced. The case of {\UNEDFONE} seems to be clear. The 
trend of the surface energy for {\UNEDFZERO} is less clean. The two groups of 
nuclei, huge and large, seem to follow slightly different slopes, and the
fit represents a compromise. The resulting surface and curvature energy have to be  taken with care.

The LDM parameters of {\UNEDFZERO} and {\UNEDFONE} are given in Table~\ref{tab:nucmat}. 
As seen in Table~I of  Ref.~\cite{[Rei06]} and Fig.~1 of Ref.~\cite{[Nik10]}, 
 symmetry coefficients  of phenomenological LDM mass models  cluster
around $a_\mathrm{sym}=30$\,MeV and $a_\mathrm{ssym}=-45$\,MeV, and  the {\UNEDFZERO} values are right in the middle. This result is not surprising, as this EDF was optimized primarily to nuclear masses.  
Indeed, the main difference between
{\UNEDFZERO} and {\UNEDFONE} lies in surface properties. Relatively low values
of $a_\mathrm{surf}$ and $a_\mathrm{ssym}$ of {\UNEDFONE} reflect the new constraints on the FI data and the neglect of the c.o.m.\ term. Again, comparing the LDM values of {\UNEDFONE} 
with those in Table~I of  Ref.~\cite{[Rei06]}, we note that the LDM parameters of {\UNEDFONE}  are closest
to those of the BSk6 EDF \cite{[Gor03]} 
($a_\mathrm{surf}=17.3$\,MeV and $a_\mathrm{ssym}=-33$\,MeV)
and the LSD LDM \cite{[Pom03]} ($a_\mathrm{surf}=17.0$\,MeV and $a_\mathrm{ssym}=-38.9$\,MeV). In Sec.~\ref{Subsec-barriers},  we shall see that the reduced surface energy of {\UNEDFONE} with respect to {\UNEDFZERO}  has profound consequences for the description of fission barriers in the actinides. 
To see this reduction more clearly, we inspect the effective 
surface coefficient
\begin{equation}
  a_{\rm surf}^{\rm (eff)} = 
 a_{\rm surf} + a_{\rm ssym} I^2.
\label{asurfeff}
\end{equation}
For $^{240}$Pu, the value of $a_{\rm surf}^{\rm (eff)}$ is 15.33\,MeV for
{\UNEDFONE}, 16.63\,MeV for {\UNEDFZERO}, 15.87\,MeV for SLy4, 15.75\,MeV for BSk6, 15.15\,MeV for SkM$^\ast$, and 15.17\,MeV for LSD.

\section{Performance of {\UNEDFONE}}
\label{Sec-results}


\subsection{Global Mass Table}
\label{Subsec-masstable}

One of the key elements required from the universal EDF is the ability to 
predict global nuclear properties, such as masses, radii, and deformations, across the nuclear chart, from drip line to drip line. We have calculated the g.s.\
mass table  with {\UNEDFONE} for  even-even nuclei with  $N,Z>8$.  
Table~\ref{table:masstableRMS} contains the rms deviations from experiment
for binding energies, separation energies, averaged three-point odd-even mass differences, and proton radii. Since the set of fit observables constraining {\UNEDFONE} is biased toward heavy nuclei, we also  show  rms deviations for 
light ($A < 80$) and heavy ($A \ge 80$) subsets.

\begin{table}[ht]
\begin{center}
\caption{Root-mean-square deviations from the experimental values for
{\UNEDFZERO} and {\UNEDFONE} for different observables calculated in
even-even systems: binding energy $E$, two-neutron separation energy $S_{2n}$,
two-proton separation energy $S_{2p}$, three-point odd-even mass difference $\tilde{\Delta}_{n}^{(3)}$ (all in MeV), and rms proton  radii $R_{\rm p}$ (in fm).
Columns are observable, RMS deviation for {\UNEDFZERO} and 
{\UNEDFONE}, and the number of data points.}
\label{table:masstableRMS}
\begin{ruledtabular}
\begin{tabular}{lrrr}
Observable & {\UNEDFZERO} & {\UNEDFONE} & \# \\
\hline
$E$                      & 1.428 & 1.912 & 555 \\
$E$ ($A<80$)             & 2.092 & 2.566 & 113 \\
$E$ ($A\ge 80$)          & 1.200 & 1.705 & 442 \\[5pt]
$S_{2\rm n}$             & 0.758 & 0.752 & 500 \\
$S_{2\rm n}$ ($A<80$)    & 1.447 & 1.161 & 99 \\
$S_{2\rm n}$ ($A\ge 80$) & 0.446 & 0.609 & 401 \\[5pt]
$S_{2\rm p}$             & 0.862 & 0.791 & 477 \\
$S_{2\rm p}$ ($A<80$)    & 1.496 & 1.264 & 96 \\
$S_{2\rm p}$ ($A\ge 80$) & 0.605 & 0.618 & 381 \\[5pt]
$\tilde{\Delta}_{n}^{(3)}$ &  0.355 & 0.358 & 442 \\
$\tilde{\Delta}_{n}^{(3)}$ ($A<80$) &  0.401 & 0.388 & 89 \\
$\tilde{\Delta}_{n}^{(3)}$ ($A\ge 80$) &  0.342 & 0.350 & 353 \\[5pt]
$\tilde{\Delta}_{p}^{(3)}$ &  0.258 & 0.261 & 395 \\
$\tilde{\Delta}_{p}^{(3)}$ ($A<80$) &  0.346 & 0.304 & 83 \\
$\tilde{\Delta}_{p}^{(3)}$ ($A\ge 80$) &  0.229 & 0.248 & 312 \\[5pt]
$R_{\rm p}$ & 0.017 & 0.017 & 49 \\
$R_{\rm p}$ ($A<80$)  & 0.022 & 0.019 & 16 \\
$R_{\rm p}$ ($A\ge 80$)  & 0.013 & 0.015 & 33 
\end{tabular}
\end{ruledtabular} 
\end{center}
\end{table}

\begin{figure}[ht]
\center
  \includegraphics[width=0.9\linewidth]{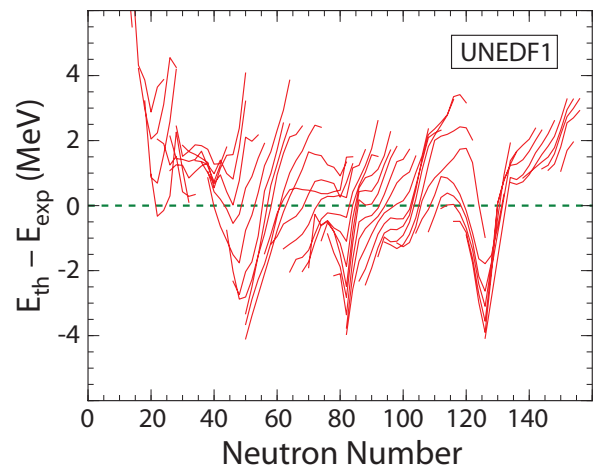}
  \caption{(Color online)  Binding energy residuals 
  between {\UNEDFONE} results and experiment for 555 even-even nuclei.
Isotopic chains of a given element are connected by lines.
}
  \label{fig:de}
\end{figure}

Figure~\ref{fig:de} displays the  binding energy residuals (i.e., deviations from experiment).  From this figure and Table~\ref{table:masstableRMS},  we can 
see a couple of trends. First, the energy residuals with {\UNEDFONE}
are  larger than those for {\UNEDFZERO}. This result is not surprising, as the new data on FI and the removal of the center-of-mass correction strongly disfavors the lightest nuclei during the optimization process. 
Second, the characteristic arc-like behavior between
the magic numbers is pronounced, although this trend is much weaker than, for example, 
for the SLy4 functional (see Fig. 7 of Ref.~\cite{[Kor10]}).

\begin{figure}[ht]
\center
  \includegraphics[width=0.8\linewidth]{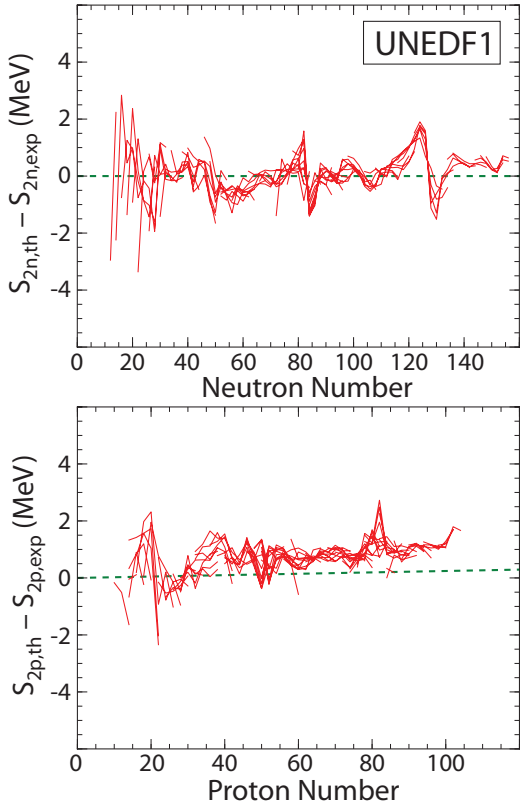}
  \caption{(Color online) Two-neutron  (top) and two-proton (bottom)
           separation energy residuals   between {\UNEDFONE} results and experiment.}
  \label{fig:ds2n}
\end{figure}

In Fig. \ref{fig:ds2n} we display the residuals of two-neutron and two-proton 
separation energies. Again, the emphasis of ${\UNEDFONE}$ on heavy 
nuclei is clearly seen, and the corresponding rms deviations in Table~\ref{table:masstableRMS} quantify this feature. Notice that two-proton separation energies are systematically overestimated. The same trend is  
observed for the {\UNEDFZERO} functional. We can speculate about 
sources for this effect: (i) Following the arguments of \cite{[Bul96]}, 
one may argue that the standard Skyrme functionals, such as {\UNEDFZERO} and {\UNEDFONE},  lack the capability to describe 
many-body Coulomb effects; (ii) The explicit contribution of the Coulomb field to the  pairing channel \cite{[Ang01],[Les08]} is not taken into account. It is expected  that separate pairing strengths for neutrons and protons, as in {\UNEDFONE}, will partly account 
for this missing contribution \cite{[Ber09a]}. However, the existence of 
nontrivial correlations between pairing strengths and other parameters 
of the functional (see Table \ref{table:correlation}) may have consequences 
for  observables such as  two-proton separation energies. 

To compare {\UNEDFZERO} and {\UNEDFONE} quantitatively, we can assess their performance on various observables listed in  Table~\ref{table:masstableRMS}. 
It is expected that since new constraints on fission isomers have been added when optimizing {\UNEDFONE} while keeping the same   number of parameters optimized, the rms deviations for masses and separation energies must increase.
Indeed,  the rms deviation for the masses is slightly worse for {\UNEDFONE}, for both light and heavy nuclei. Interestingly, the 
quality of  $S_{2\rm n}$ values remains roughly the same in both cases, as is true also for odd-even mass differences and proton radii.


\subsection{Spherical Shell Structure}
\label{Subsec-shell}

The nuclear shell structure has a substantial impact on many nuclear properties. Notably, the single-particle levels close to the Fermi surface affect many nuclear properties such as the strength of pairing correlations and deformability. Compared with our previous work \cite{[Kor10]}, the s.p.\ energies that we report here have been obtained from proper blocking calculations at the HFB+LN level \cite{[Sch10]}, instead of being the eigenvalues of the HF Hamiltonian. This choice is motivated by the need to stay within a logically consistent framework: both {\UNEDFZERO} and {\UNEDFONE} have been optimized at the HFB+LN level, and hence should be employed exclusively in this context. Moreover, in the nuclear mean-field theory with effective interactions, HF eigenvalues are a poor representation of s.p.\ energies; see \cite{[Col10],[Lit07]} for a recent study. In a  DFT approach, however, it is assumed that the generalized form of the energy density may effectively account for beyond mean-field effects such as particle-vibration couplings. In addition to this  theoretical argument, let us recall that s.p.\ energies are not observables but model-dependent quantities extracted experimentally from binding energies of excited states in odd nuclei. Systematic errors  can thus be reduced by working exclusively with binding energies.

\begin{figure}[ht]
\center
  \includegraphics[width=\linewidth]{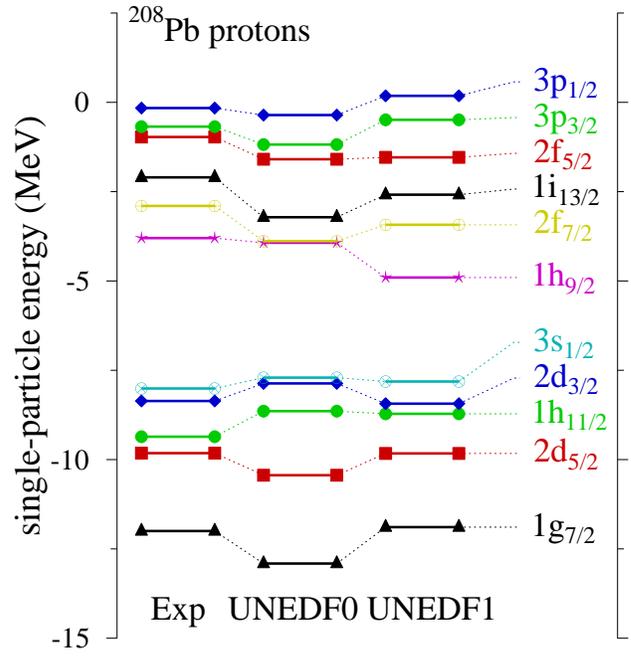}
  \caption{(Color online) Single-proton energies for $^{208}$Pb, 
           calculated with {\UNEDFZERO} and {\UNEDFONE} EDFs compared to the experimental values of Ref.~\cite{[Sch07]}.}
  \label{fig:spe208p}
\end{figure}

To this end, we computed a number of one-quasi-particle (q.p.) configurations for the odd-mass neighbors of $^{16}$O, $^{40}$Ca, $^{48}$Ca, $^{56}$Ni, $^{132}$Sn, and $^{208}$Pb. Calculations were done at the equal filling approximation, which is an excellent approximation to the full time-reversal, symmetry-breaking blocking scheme \cite{[Sch10]}. Blocking q.p. states induces a small shape polarization \cite{[Zal08]}, which in turn leads to a fragmentation of spherical s.p.\ orbitals of angular momentum $j$ into $2j+1$ levels $\Omega = -j, \dots, +j$. In principle, the ``experimental'' s.p.\ energy should be the average energy over all the $2j+1$ blocking configurations. However, a state with the projection $\Omega$ can belong to any spherical orbit $j\geq |\Omega|$, which could potentially complicate the identification for low-$\Omega$ values. We therefore associate a spherical orbit with spin $j$ to the blocking configuration with the maximum projection $\Omega = +j$. We have verified that the amount of splitting in all these nuclei does not exceed 150 keV: for the proton $i_{13/2}$ orbit in $^{208}$Pb, it is 77 keV, and for the neutron 1$f_{7/2}$ state in $^{40}$Ca, it is 124 keV. This implies that the error induced by cherry-picking a single state out of a $2j+1$ nondegenerate candidate instead of the average is only on the order of 60 keV.

Figure~\ref{fig:spe208p} compares  single-proton  
energies in $^{208}$Pb computed with {\UNEDFZERO} and {\UNEDFONE} with the experimental values. Although the differences between these two 
functionals are small, we note that {\UNEDFONE} 
improves slightly the description of several high-$j$ levels except for a too low position of the 1$h_{9/2}$ proton orbit that reduces the size of the $Z$=82 proton gap. The situation is  similar 
for the neutron single-particle energies in $^{208}$Pb. For both {\UNEDFZERO} 
and {\UNEDFONE} the effective mass is close to one, which probably explains the 
fairly good reproduction of the level density in $^{208}$Pb. In the lighter doubly magic nuclei, 
differences between {\UNEDFZERO} and {\UNEDFONE} s.p.\ energies 
are somewhat larger. The magic gaps in Ca isotopes are now better reproduced, although the $N=28$ gap in $^{48}$Ca is still too low.


\subsection{Superdeformed States and Fission Barriers}
\label{Subsec-barriers}

Table \ref{table:sd} lists the excitation energies of superdeformed (SD) fission isomers in the actinide region and SD  bandheads in the mass $A\sim 190$ region calculated with the {\UNEDFZERO} and {\UNEDFONE} parameterizations. Contrary to fission isomers, SD bandheads in neutron-deficient lead and mercury isotopes were not included in the objective function, since the prolate-oblate shape coexistence effects, not  captured by current functionals, are well known in these nuclei \cite{[Woo92],[Naz93a]}. Indeed, calculations with  {\UNEDFONE} predict an oblate ground state at $\beta$=$-0.2$ -- $-0.15$ in all the Hg-Pb isotopes considered, coexisting with a slightly higher spherical minimum. By contrast, the ground state of the three lead isotopes is spherical with {\UNEDFZERO}. The fact that the spherical configuration
is disfavored in these nuclei can be traced back to a too-low $Z$=82 spherical proton gap in {\UNEDFONE},  (Sec.~\ref{Subsec-shell}).

\begin{table}[ht]
\begin{center}
\caption{Excitation energies (in MeV) of  fission isomers in the actinides and superdeformed bandheads in the neutron-deficient Hg and Pb nuclei calculated with {\HFBTHO}. The values predicted with {\UNEDFZERO} and {\UNEDFONE} are compared with experiment.}
\label{table:sd}
\begin{ruledtabular}
\begin{tabular}{rlllr}
Nucleus & {\UNEDFZERO} & {\UNEDFONE} &  Exp. & Ref. \\
\hline\\[-7pt]
$^{236}$U   & 5.28 & 2.42  & 2.75  & \cite{[Sin02]} \\
$^{238}$U   & 5.73 & 2.71  & 2.557 & \cite{[Sin02]} \\
$^{240}$Pu  & 5.74 & 2.51  & 2.8   & \cite{[Sin02]} \\
$^{242}$Cm  & 5.27 & 1.85  & 1.9   & \cite{[Sin02]} \\[5pt]
$^{192}$Hg  & 6.33 & 2.62  & 5.3   & \cite{[Lop96]} \\
$^{194}$Hg  & 7.27 & 3.79  & 6.017 & \cite{[Hen94s]} \\
$^{192}$Pb  & 5.20 & 1.25  & 4.011 & \cite{[Hen91]} \\
$^{194}$Pb  & 5.99 & 1.99  & 4.643 & \cite{[Hau97]} \\
$^{196}$Pb  & 7.26 & 3.52  & 5.63  & \cite{[Wil05]} \\
\end{tabular}
\end{ruledtabular}
\end{center}
\end{table}

All values listed in Table~\ref{table:sd} were obtained with the {\HFBTHO} code using 
the  same large HO basis as used for the optimization. In particular, the 
deformation of the basis was spherical for the ground state, and was deformed with 
$\beta_{2} =0.4$ for the FI; see Sec. \ref{Subsec-precision}.
As can be seen from Table~\ref{table:sd}, the optimization improves dramatically the rms deviation for  the actinide nuclei included in the fit, going from 3.02~MeV in {\UNEDFZERO} to 0.23~MeV in {\UNEDFONE}. At the same time, the optimization deteriorates the description of SD excitations in the Hg and Pb isotopes. To understand this behavior, we again compute the effective surface coefficient $a_{\rm surf}^{\rm (eff)}$ (\ref{asurfeff}) for $^{194}$Pb. It is particularly low, 16.0\,MeV, for {\UNEDFONE}. Indeed, it is  17.64\,MeV for {\UNEDFZERO}, 17.1\,MeV for SLy4, 16.50\,MeV for BSk6, and 16.36\,MeV for SkM$^\ast$. In addition, the reduced $Z$=82 magic gap in {\UNEDFONE} energetically favors deformed  and SD states. Consequently, both bulk energy and shell effects of {\UNEDFONE} conspire to reduce the excitation energy of SD states in the Pb isotopes. In view of the major shape coexistence 
effects recalled earlier, this behavior is not too worrisome.
\begin{table*}[ht]
\begin{center}
\caption{Empirical 
 and theoretical inner barrier heights $E_A$ (in MeV) for selected actinide nuclei. The rms deviations from experiment are $\Delta E_A$ are shown in the last row.}
\label{firstbarrierheights}
\begin{ruledtabular}
\begin{tabular}{cccccccc}
Nuclide & Exp. \cite{[Cap09]} & HFB+Fit \cite{[Gor09a]} & ETFSI \cite{[Mam01]} & FRLDM \cite{[Mol09]} & HFB-14 \cite{[Gor07]} & SkM$^\ast$ & {\UNEDFONE}\\
\hline\\[-7pt] 
$^{236}$U & 5.00 & 5.52 & 5.20 & 4.45 & 5.52 & 6.93 & 6.39\\
$^{238}$U & 6.30 & 5.80 & 5.70 & 5.08 & 5.93 & 7.25 & 6.50\\
$^{238}$Pu & 5.60 & 5.57& 5.40 & 5.27 & 5.96 & 7.39 & 6.83\\
$^{240}$Pu & 6.05 & 5.89 & 5.80 & 5.99 & 6.49 & 7.51 & 6.77\\
$^{242}$Pu & 5.85& 6.02 & 6.20 & 6.42 & 6.81 & 7.44 & 6.59\\
$^{244}$Pu & 5.70 & - & 6.40 & 6.59 & 6.85 & 7.82 & 6.10\\
$^{242}$Cm & 6.65 & 6.20 & 6.10 & 6.56 & 6.75 & 8.76 & 7.12\\
$^{244}$Cm & 6.18  & 6.18 & 6.40 & 6.92 & 7.10 & 8.81 & 6.99\\
$^{246}$Cm & 6.00 &  6.00 & 6.50 & 7.01 & 7.31 & 8.41 & 6.69\\
$^{248}$Cm & 5.80 & - & 6.50 &  6.80 & 7.25 & 7.94 & 6.12 \\[3pt]
$\Delta E_A$ & & & 0.47 & 0.75 & 0.87 & 1.97 & 0.79 
\end{tabular}
\end{ruledtabular}
\end{center}
\end{table*}

\begin{table*}[ht]
\begin{center}
\caption{Similar to Table~\ref{firstbarrierheights} except for the outer barrier heights   $E_B$ (in MeV).}
\label{secondbarrierheights}
\begin{ruledtabular}
\begin{tabular}{cccccccc}
Nuclide & Exp. \cite{[Cap09]}  & HFB+Fit \cite{[Gor09a]} & ETFSI \cite{[Mam01]} & FRLDM  \cite{[Mol09]} & HFB-14 \cite{[Gor07]} & SkM$^\ast$ & {\UNEDFONE}\\
\hline\\[-7pt] 
$^{236}$U & 5.67 & 6.03 & 4.00 & 5.03 & 6.03 & 6.70 & 5.56 \\
$^{238}$U & 5.50 & 6.17 & 4.90 & 5.64 & 6.48 & 7.36 & 6.42 \\
$^{238}$Pu & 5.10 & 5.35 & 2.90 & 4.47 & 5.24 & 5.99 & 4.62 \\
$^{240}$Pu & 5.15 & 5.73 & 3.40 & 4.91 & 5.61 & 6.40 & 5.42 \\
$^{242}$Pu & 5.05 & 5.61 & 3.60 & 5.72 & 6.02 & 6.90 & 6.20 \\
$^{244}$Pu & 4.85 & - & 3.90 & 6.47 & 6.25 & 7.49 & 6.50 \\
$^{242}$Cm & 5.00 & 4.90 & 1.70 & 4.45 & 4.51 & 6.31 & 4.08 \\
$^{244}$Cm & 5.10 & 5.10 & 2.10 & 5.07 & 4.83 & 7.00 & 5.03 \\
$^{246}$Cm & 4.80 & 4.80 & 2.40 & 5.87 & 5.23 & 7.42 & 5.51 \\
$^{248}$Cm & 4.80 & - & 2.60 & 6.65 & 5.25 & 7.32 & 5.55\\[3pt]
$\Delta E_B$ & & & 2.11 & 0.94 & 0.70 & 1.89 & 0.84 
\end{tabular}
\end{ruledtabular}
\end{center}
\end{table*}
In the $A\sim  190$ region, the experimental
uncertainty of the SD bandhead comes from the extrapolation
of the rotational band down to spin $0^+$. The associated error is estimated to be very small, around  5 keV. In the actinides, 
experimental excitation energies of FI are usually determined with larger uncertainties. While the experimental error bar 
is only about 5-10 keV for $^{236,238}$U, it grows to  about 200 keV for $^{240}$Pu and $^{242}$Cm. For  the fission isomer of $^{240}$Pu, recent measurements 
lower its excitation energy by about 500\,keV to  roughly 2.25\,MeV \cite{[Hun01]} compared with the standard value \cite{[Sin02]} adopted in this work. 
Because of the relatively large uncertainty, $w_i=0.5$\,MeV, adopted for FI energies in our  objective function, these experimental uncertainties are not going to significantly alter the final optimization. For  future work, however, better experimental determination of FI bandheads should become a priority.

Our long-term goal is to develop an 
EDF that can accurately predict and describe fission observables in heavy and superheavy nuclei. We present here some results of  spontaneous fission pathway calculations  with {\UNEDFONE}. 
All fission calculations were done with version 2.49t of the  
code {\HFODD} \cite{[Sch11]} that can break all self-consistent symmetries along the fission path. At each point along the collective trajectory, the HO 
basis deformation and frequencies are determined from a standard nuclear surface, parametrized by surface deformations $\alpha_{\lambda\mu}$.  The  deformations  were chosen to minimize the energy as a function of the requested quadrupole moment $\overline{Q}_{20}$, 
according to the following expressions:
\begin{equation}
\begin{array}{rcl}
\alpha_{20} & = & A~\overline{Q}_{20}^3 + B~\overline{Q}_{20}^2 + C~\overline{Q}_{20}, \\
\alpha_{40} & = & 0.01, 
\end{array}
\end{equation}
with  $A = 3.16721\times 10^{-8}$\,b$^{-3}$, 
$B =  -2.75505 \times 10^{-5}$\,b$^{-2}$, 
$C = 0.00954925$\,b$^{-1}$, and all remaining deformations  $\alpha_{\lambda\mu}=0$.
The basis contains up to $N = 31$ shells and up to $N = 1140$ 
states. Such an extended basis  was previously applied in the systematic study of fission 
barriers in the transfermium  region and yielded excellent convergence \cite{[Sta05a],[Sta09]}. We have checked that {\HFODD}, with a reduced HO basis as compared with {\HFBTHO} calculations, reproduces  the  {\HFBTHO} energies of FI bandheads in Table~\ref{table:sd} with an accuracy of 100-200\,keV. We  consider this agreement satisfactory considering other uncertainties involved.

\begin{figure}[ht]
\center
  \includegraphics[width=\linewidth]{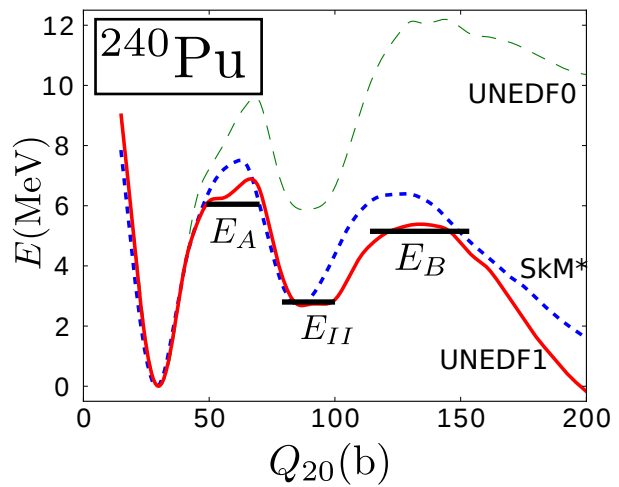}
  \caption{(Color online) Fission pathway for $^{240}$Pu along the mass quadrupole moment $Q_{20}$ calculated using {\HFODD}   with
           SkM$^\ast$, {\UNEDFZERO}, and {\UNEDFONE} EDFs.
           The ground state energies have been normalized to zero. $E_{II}$,
           $E_{A}$,  and $E_{B}$ indicate, respectively,  the experimental energy of fission isomer and
            the inner and outer barrier heights \cite{[Cap09]}.}
  \label{fig:fission}
\end{figure}

As an example, Fig.~\ref{fig:fission}  displays
the potential energy curve of $^{240}$Pu as a function of the mass quadrupole moment $Q_{20}$. Triaxiality and reflection asymmetry 
effects are included for all calculations.  The large-deformation behavior of the potential energy curve obtained by {\UNEDFZERO} is 
typical in this region, so that the outer barrier heights obtained by {\UNEDFZERO} are systemmatically much higher than empirical values. We note that the {\UNEDFONE} functional yields both the inner and outer barrier  in $^{240}$Pu fairly close to experiment. Both {\UNEDFZERO} and {\UNEDFONE} functionals also yield the g.s.\ binding energy that is close to the empirical value. 

The {\UNEDFONE} results for fission barrier heights  in selected actinide nuclei are listed in Tables~\ref{firstbarrierheights} (inner barrier) and \ref{secondbarrierheights} (outer barrier).
For comparison, we also list the empirical barrier heights from
the Reference Input Parameter Library (RIPL-3) \cite{[Cap09]}; the HFB fission barriers obtained by fitting the neutron-induced fission cross section
\cite{[Gor09a]}; and predictions of 
 ETFSI \cite{[Mam01]}, FRLDM  \cite{[Mol09]}, and HFB-14 \cite{[Gor07]} models, together with {\HFODD} calculations with the SkM$^\ast$ EDF.
(For  SkM$^\ast$  predictions including the energy correction due to the rotational
zero-point motion, see Ref.~\cite{[Bon04]}.)
Overall, the description of experimental data by {\UNEDFONE} is very reasonable, with the rms deviations from experimental values of $E_A$ and $E_B$  comparable to the values obtained  in more phenomenological models. 
One can thus conclude that  fission barriers are reliably described at the HFB level with the {\UNEDFONE} functional. This result is remarkable since it was obtained by adding only four excitation energies to the dataset.

There seems to exist an interesting relation between barrier heights and the surface thickness. We have evaluated the surface thickness in $^{208}$Pb from the charge form factor $\sigma_\mathrm{ch}$ as defined, e.g., in Refs.~\cite{[Ben03],[Klu09]} and found $\sigma_\mathrm{ch}=0.932$\,fm for {\UNEDFZERO}  and $\sigma_\mathrm{ch}=0.907$\,fm for {\UNEDFONE}. This is to be compared with the measured value of $\sigma_\mathrm{ch}=0.913$\,fm \cite{[Klu09]}. It is apparent that the  EDF which does well on fission barriers  also performs well for surface thickness. The functionals SV-min and SV-bas which included $\sigma_\mathrm{ch}$ in the fit, and yield values around 0.91 fm  for this quantity, happen to perform  well concerning barrier heights \cite{[Erl10]}. The relation between fission barriers and surface thickness  deserves closer inspection in future work.


\subsection{Neutron Drops}

Recently, there has been a  considerable interest in studies of inhomogeneous neutron matter by considering finite systems of $N$ neutrons; specifically, neutron drops \cite{[Pud96],[Gan11]}. Since neutron drops are not self-bound \cite{[Pie03]}, an external potential must be used to confine them.  
By studying  neutron drops, one can test different ab initio approaches and their correspondence to DFT calculations \cite{[Pud96],[Gan11]}; investigate the validity of the density matrix expansion \cite{[Bog11]}; and develop a theoretical  link between  neutron-rich nuclei and the  neutron matter found in the neutron star crust \cite{[Rav83]}.
\begin{figure}[ht]
\center
  \includegraphics[width=1.0\linewidth]{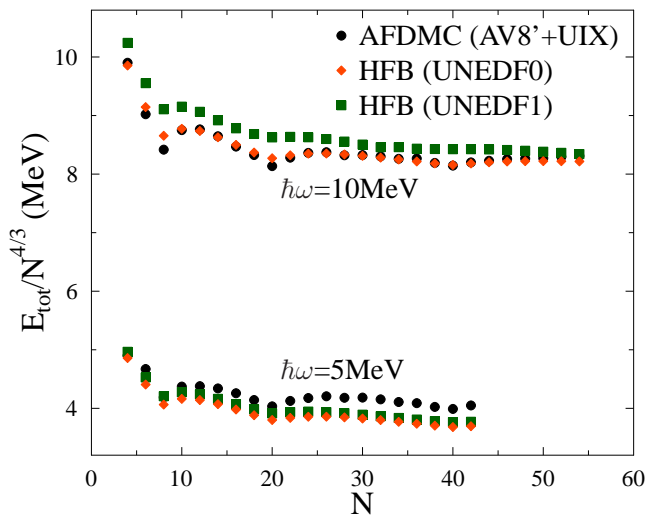}
  \caption{(Color online) Comparison of {\UNEDFZERO} and {\UNEDFONE} predictions for  the energy  $N$-neutron drops trapped in a HO potential with $\hbar\omega=5$\,MeV and 10\,MeV with  the AFDMC ab initio results of Ref.~\cite{[Gan11]}.}
  \label{fig:ndroplets}
\end{figure}

Figure \ref{fig:ndroplets} presents the results of {\UNEDFZERO} and {\UNEDFONE} calculations for neutron drops 
confined by two external HO traps with $\hbar\omega=5$\,MeV and 10\,
MeV. The DFT results are compared with 
ab initio AFDMC benchmark calculations 
of Ref.~\cite{[Gan11]} employing the  AV8' nucleon-nucleon 
and Urbana IX three-nucleon force. As can 
be seen, {\UNEDFZERO} reproduces AFDMC results well, especially 
since the functional has not been constrained to finite neutron matter. The agreement with {\UNEDFONE} calculations is 
also  good, especially for a  softer trap with $\hbar\omega=5$\,MeV.

In future EDF optimizations we shall include ab initio predictions for neutron 
drops into the dataset. By providing unique constraints on finite neutron matter, such pseudo-data are expected to improve the description of very neutron-rich 
nuclei and diluted neutron matter. The results shown in Fig.~\ref{fig:ndroplets}
indicate that {\UNEDFZERO} and {\UNEDFONE} functionals represent excellent starting points for such optimizations.


\section{Conclusions}
\label{Sec-conclusions}

By performing nuclear energy density optimization at the  deformed HFB level, we have arrived at the new Skyrme parameterization {\UNEDFONE}. Our main focus was to improve the description of fission properties of the actinide nuclei and to provide a high-quality functional for time-dependent applications involving heavy systems. The only notable change in the form of the energy density as compared with our previous work \cite{[Kor10]} was the removal of the center-of-mass correction. For the $\chi^{2}$-minimization, we used the derivative-free {\algo} algorithm. Compared with  {\UNEDFZERO}, the dataset was enlarged by adding ground-state masses of three deformed actinide nuclei and  excitation energies of  fission isomers in  $^{236,238}$U, $^{240}$Pu, and $^{242}$Cm. 
For the optimal parameter set, we carried out a sensitivity analysis to obtain information about the standard deviations and correlations among the parameters. We conclude that {\UNEDFONE} remains as robust under a change of individual data as {\UNEDFZERO}. 

Overall, {\UNEDFONE} provides a description of global nuclear properties that is almost as good  as that of  {\UNEDFZERO}. 
Not  surprising, the quality of data reproduction is  slightly degraded: by adding a new type of data (fission isomers), one is likely to worsen the fit for other observables. The most striking feature of {\UNEDFONE} is its ability to reproduce 
the empirical fission barriers in the actinide region. We find it encouraging that, by including only a handful 
of fission isomer bandheads, deformation properties of the functional seem  
well constrained. 
Another unanticipated property of {\UNEDFZERO} and {\UNEDFONE} is their ability to reproduce ab initio results for trapped neutron drops. This is significant because such pseudo-data will be used in future EDF optimizations. 

In addition to imposing new constraints on neutron drops, in the next step we intend to improve the spectroscopic quality of {\UNEDF} functionals by considering the experimental data on spin-orbit splittings and shell gaps. We shall also improve the density dependence of the kinetic term by adding new constraints on giant resonances. Meanwhile, the functional {\UNEDFONE} developed in this work will be the  input of choice  for  microscopic studies of the nuclear fission process.

\bigskip
\begin{acknowledgments}
We are grateful to A. Staszczak and J.~Mor\'e for helpful discussions. 
This work was 
supported by the U.S. Department of Energy under 
Contract Nos.\ DE-FC02-09ER41583 (UNEDF SciDAC Collaboration), DE-FG02-96ER40963 
 (University of Tennessee), 
DE-FG52-09NA29461 (the Stewardship Science Academic Alliances program),   DE-AC07-05ID14517 (NEUP grant sub award 00091100), and DE-AC0Z-06CA11357 (Argonne National Laboratory), and was 
partly performed under the auspices of the U.S.\ Department of Energy by Lawrence 
Livermore National Laboratory under Contract DE-AC52-07NA27344. Funding was also
provided by the U.S.\ Department of Energy Office of Science, Nuclear 
Physics Program pursuant to Contract DE-AC52-07NA27344 Clause B-9999, Clause 
H-9999, and the American Recovery and Reinvestment Act, Pub. L. 111-5. 
Computational resources were provided through an INCITE award ``Computational 
Nuclear Structure'' by the National Center for Computational Sciences (NCCS) and 
National Institute for Computational Sciences (NICS) at Oak Ridge National 
Laboratory, and through an award by the Laboratory Computing Resource Center at 
Argonne National Laboratory.

\end{acknowledgments}



\end{document}